\documentclass[superscriptaddress,showpacs,aps,twocolumn,floatfix,prl,longbibliography]{revtex4-1}
\usepackage{bm,amsmath,amssymb}
\usepackage{amssymb}
\usepackage{amsmath}
\usepackage{graphicx,bm}
\usepackage{verbatim}
\usepackage[dvipsnames]{xcolor}
\usepackage{soul}
\usepackage{ bbold }
\usepackage{xspace}
\usepackage{tikz}
\usetikzlibrary{snakes, arrows, backgrounds, calc, fadings, decorations.pathreplacing, shadings, through, intersections, angles, shapes.geometric}

\usepackage{pgfmath}
\usepackage{pgfplots}
 
\pgfplotsset{compat = newest}

\usepackage[colorlinks]{hyperref}
\AtBeginDocument{%
 \hypersetup{ 
 linkcolor=blue,
 filecolor=magenta, 
 urlcolor=blue,
 citecolor=blue,
 colorlinks=true,
 }
}

\usepackage[T1]{fontenc}
\usepackage[utf8]{inputenc}

\newcommand{\p}	{\partial}
\newcommand{\bp}{\bar{\partial}}

\newcommand{\R}	{\mathbb{R}}

\newcommand{\cD}{\mlo}

\newcommand{\cM}{\mathcal{M}}
\newcommand{\cN}{\mathcal{N}}
\newcommand{\cO}{\mathcal{O}}

\DeclareMathOperator{\tr}{\tr}

\newcommand{\corr}[1] {\left\langle{#1}\right\rangle}

\newcommand{\bz} {\bar{z}}
\newcommand{\bw} {\bar{w}}
\newcommand{\bphi} {\bar{\phi}}

\newcommand{\bJ} {\bar{J}}

\newcommand{\bh} {\bar{h}}

\newcommand{\bT} {\bar{T}}
\newcommand{\bg} {\bar{g}}

\newcommand{\bs} {\bar{s}}

\renewcommand{\tr}	{\mathrm{tr}}

\newcommand\id		{\mathbf{1}}

\newcommand{\ca} {\mathsf{c}}
\newcommand{\bca} {\bar{\ca}}
\newcommand{\rhbar} {\textcolor{black}{\hbar}}
\renewcommand{\st}    {\xi}
\newcommand{\Kel}     {\mathsf{K}}
\newcommand{\temp}  {\mathsf{T}}
\newcommand{\bbeta} {\beta_{\text{av}}}
\newcommand{\mlo}   {M^2_{loc}}

\newcommand\quotes[1] {``{#1}"}

\newcommand\figref[1]	{figure~\ref{#1}\xspace}

\newcommand\cn[1]	{\textcolor{teal}{(CN: {#1})}}

\begin{document}

\title{Conformal anomaly in magnetic finite temperature response of strongly interacting one-dimensional spin systems}

\author{C. Northe$^*$}
\affiliation{Institute for Theoretical Physics and Astrophysics,	Julius-Maximilians-Universit{\"a}t W{\"u}rzburg, 97074 W{\"u}rzburg, Germany}
\affiliation{Würzburg-Dresden Cluster of Excellence ct.qmat}
\thanks{These two authors contributed equally to this work.}	

\author{C. Zhang$^*$}
\affiliation{Institute for Theoretical Physics and Astrophysics,	Julius-Maximilians-Universit{\"a}t W{\"u}rzburg, 97074 W{\"u}rzburg, Germany}
\affiliation{Würzburg-Dresden Cluster of Excellence ct.qmat}
\thanks{These two authors contributed equally to this work.}

\author{R. Wawrzy\'{n}czak}
\affiliation{Max Planck Institute for Chemical Physics of Solids, Nöthnitzer Straße 40, 01187 Dresden, Germany}

\author{J. Gooth}
\affiliation{Würzburg-Dresden Cluster of Excellence ct.qmat}
\affiliation{Max Planck Institute for Chemical Physics of Solids, Nöthnitzer Straße 40, 01187 Dresden, Germany}

\author{S. Galeski}
\affiliation{Max Planck Institute for Chemical Physics of Solids, Nöthnitzer Straße 40, 01187 Dresden, Germany}

\author{E. M. Hankiewicz}
\affiliation{Institute for Theoretical Physics and Astrophysics,	Julius-Maximilians-Universit{\"a}t W{\"u}rzburg, 97074 W{\"u}rzburg, Germany}
\affiliation{Würzburg-Dresden Cluster of Excellence ct.qmat}


\begin{abstract}
The conformal anomaly indicates the breaking of conformal symmetry (angle-preserving transformations)  in the quantum theory by quantum fluctuations and is a close cousin of the gravitational anomaly.
We show, for the first time, that the conformal anomaly controls the variance of the local magnetization $M_{loc}$ at finite temperatures in spin chains and spin ladders. This effect is perceived at constant and variable temperature across the sample.
The change of $M_{loc}$ induced by the conformal anomaly is of the order of 3-5\% of the maximal spin at one Kelvin for DIMPY or CuPzN and increases linearly with temperature.
Further, for a temperature gradient of 10\% across the sample, the time-relaxation of the non-equilibrium $M_{loc}$ is of the order of nanoseconds. 
Thus, we believe that experimental techniques such as neutron scattering, nuclear magnetic resonance~(NMR), spin noise and
ultrafast laser pumping should pinpoint the presence of the conformal anomaly. 
Therefore, we pave the road to detect the conformal anomaly in spin observables of strongly interacting low-dimensional magnets. 
\end{abstract}

\maketitle
\textit{Introduction}.---
The study of quantum anomalies is one of the most rapidly developing fields at the interface between condensed matter experiment and high-energy theory. A quantum anomaly is the violation of classical conservation laws due to quantum fluctuations. Understanding such conservation law violations has played an important role in the development of the standard model. However, experimental access to these phenomena is often hampered by the extremely high energies required in particle physics experiments. Recently, such anomalies have been accessed and experimentally studied in Weyl and Dirac semimetals. This new direction in the study of solids resulted in recent observations of signatures of the chiral ~\cite{nielsen1983adler, sonchiral, dubvcek2015weyl, lu2015weak, zhang2016signatures, zhang2017evolution} and mixed axial-gravitational ~\cite{gooth2017experimental}  anomaly manifesting as additional field-dependent contributions to the longitudal components of the  conductance and  thermoelectric conductance tensors. 

Thus far experimental studies of quantum anomalies in solids have mainly focused on signatures of anomalies in nearly free electron systems such as the Weyl semimetals. 
However, anomalies also play a role in strongly correlated systems, where interactions cannot be treated as a simple perturbation to the physics of free quasi-particles.
In this Letter, we employ quantum field theory to predict experimental signatures of the conformal anomaly in the context of strongly interacting 1D Heisenberg spin-1/2 chains and ladders. In particular, we discuss the magnitude of the effects and potential experimental tools to access the signatures of the anomalies using example parameters of two excellent material realizations of the mentioned Hamiltonians, DIMPY and CuPzN.


Recent studies of one-dimensional ($1d$) quantum magnets have shown that the ground state of both the Heisenberg spin-1/2 chain and the magnetized spin-1/2 ladder belong to the universality class of the Tomonaga Luttinger Liquid (TLL)~\cite{senechal2004introduction, Giamarchi, affleck1989houches}. The TLL, due to its one-dimensionality and divergence of susceptibilities, lives at the verge of an ordering instability making it a realization of the $z=1$ quantum critical state. As such the TLL is expected to exhibit no internal energy scales apart from temperature and is thus expected to feature the conformal anomaly. Indeed the free energy of quantum critical systems \cite{bonner1988quantum, affleck1988universal} was discussed in the context of the conformal anomaly, however those works focus on the appearance of the conformal anomaly in thermal properties such as specific heat~\cite{affleck1988universal} and thermal conductivity \cite{heidrich2002thermal}. In addition, thermal effects for curved space were connected to the conformal anomaly in  ~\cite{ langmann, stone2012gravitational, bermond2022anomalous}, which are difficult to observe in experiments. Altogether, to date there exist no predictions on how the conformal anomaly is manifested in spin observables.  
Rapid development of spectroscopic techniques and growth of excellent material realizations of low-dimensional magnetic Hamiltonians~\cite{bouillot2011statics, schmidiger2013spectrum, blosser2018z, galeski2022t, wu2019tomonaga, nikitin2021multiple, breunig} has opened a new avenue for observation of signatures of the conformal anomaly in flat space time by directly investigating spin observables.  

In the following, we explore this new direction of research and demonstrate how signatures of the conformal anomaly appear in the variance of local magnetization at finite temperature as well as in dynamics of variance of local magnetization upon imposing temperature gradients.  
We predict measurable effects thereof in neutron scattering, NMR, atomic force microscopy (AFM) and spin noise. Moreover, we predict that the dynamics of  the local magnetization should lie in the range of nanoseconds, and thus accessible in ultrafast laser spectroscopy.

\begin{figure}
\includegraphics[scale=.32]{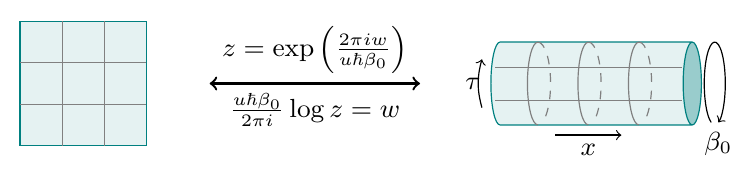}
\caption{Conformal transformations are angle-preserving, as indicated by the grids. The transformation here relates a zero-temperature system, drawn as plane, to a thermal ensemble at temperature $\temp$, depicted by a cylinder of circumference $\beta_0=(k_B\temp)^{-1}$. }
\label{figConfTransform}
\end{figure}
\textit{The conformal anomaly}.---
In order to detect the presence of the conformal anomaly in spin observables, it is first necessary to understand where the anomaly resides. Conformal transformations are angle-preserving transformations as seen in \figref{figConfTransform}. For conformal symmetry, Noether's theorem implies the tracelessness of the energy-momentum tensor. Quantum mechanically, this is altered by the conformal anomaly: the trace of the energy-momentum tensor becomes proportional to the conformal anomaly $\ca$ itself and the Ricci scalar of the manifold $\cM$ the system resides on. The latter is a measure of the curvature of $\cM$ and it vanishes in flat spacetime. Thus, this way of accessing the conformal anomaly is not optimal for experiments. However, as we review here briefly, the conformal anomaly $\ca$ appears in flat space too. One place to look for it is in the expectation value of the left-moving and right-moving (component of the) energy-momentum tensor, which vanish in the ground state, $\corr{T}=0=\corr{\bT}$. However, upon exciting the system thermally up to temperature $k_B\temp=\beta_0^{-1}$, the quantum system responds, via the conformal anomaly. The quantum system lives then on a cylinder with spatial coordinate $x\in\R$ and compact Euclidean time coordinate $\tau\in[0,\hbar\beta_0]$. It is convenient to parameterize this thermal cylinder with complex coordinates $w=u\tau-ix$ and $\bw=u\tau+ix$, where $u$ is the Fermi velocity. The thermal cylinder is connected to the zero temperature setup, the latter being thought of simply as the complex plane, using the conformal transformation, $ z=\exp\left(\frac{2\pi i w}{u\hbar\beta_0}\right)$ and its inverse $ w=\frac{u\hbar\beta_0}{2\pi i}\log z\,,$
and similarly for $\bw \leftrightarrow \bz$, see \figref{figConfTransform}. As reviewed in \cite{appendix}, this leads to a shift in energy, which is controlled by the conformal anomaly,
\begin{align}\label{TcylinderOnePt}
 \corr{T(w)}_{\beta_0}=
 \frac{\pi^2\ca}{6u\hbar\beta_0^2},
 \quad
 \corr{\bT(\bw)}_{\beta_0}=
 \frac{\pi^2\bca}{6u\hbar\beta_0^2}\,.
\end{align}
A priori $\bca\neq\ca$, however, in the systems of interest here the conformal anomalies coincide, $\ca=\bca$.


\textit{Spin Systems}.---
We consider $1d$ spin system of length $L$, such as 
a spin-$\frac{1}{2}$ chain with Hamiltonian
\begin{align}
 H
 &=\sum_i\left[J\sum_{b=x,y}S^b_iS^b_{i+1}+J_zS^z_iS^z_{i+1}-BS^z_i\right]\label{SpinChain}
 %
\end{align}
where $B$ is a magnetic field perpendicular to the chain. If $J_z=J_\perp$, the Hamiltonian \eqref{SpinChain} becomes that of the Heisenberg chain
, 
$H_{HC}=J\sum_i(\vec{S}_i\cdot\vec{S}_i-BS^z_i)$. 
The spin-$z$ operator bosonizes according to~\cite{Giamarchi, affleck1989houches}
\begin{align}
 %
 \frac{S^z(x)}{a\,\hbar}&=m\,\id-\frac{\p_x\varphi}{2\pi R}+\frac{1}{\pi}\cos\left(\frac{\varphi(x)}{R}-2k_F(h)x\right)\label{SzBoso}
\end{align}
where $a$ is the lattice constant, the compactification radius $R=\frac{1}{\sqrt{4\pi K}}$ is related to the Luttinger parameter $K$, $u$ is the Fermi velocity and $k_F(h)=k_F+\pi m$ is the Fermi momentum shifted by the magnetization per site, $m=\corr{S^z(x)}/L$. In this paper we are not concerned with the ladder operators $S^\pm$, and thus we omit their bosonization prescription. The spin chain Hamiltonian \eqref{SpinChain} reduces to that of the TLL \cite{affleck1989houches, Giamarchi}, $H_{TLL}=\frac{\hbar u}{2}\int dx\left[(\p_x\varphi)^2/K+K(\p_x\theta)^2\right]$
up to terms irrelevant in the renormalization group sense. This system is classically conformally invariant. However, upon quantization it features the conformal anomaly, which we exploit below.

A more elaborate case of a $1d$ spin system is a spin ladder, in which two spin-$\frac{1}{2}$ Heisenberg chains, i.e. with $J_z=J$ in their Hamiltonian \eqref{SpinChain}, are coupled to each other via an interaction term $H_{ladder}=H_{HC}^{(1)}+H_{HC}^{(2)}+J_{\perp}\sum_i \vec{S}^{(1)}_i\cdot\vec{S}^{(2)}_i$.
The superscripts indicate the respective chain. In the low-energy limit and above a critical magnetic field $B_{\text{crit}}$, the spin ladder is also described by the TLL Hamiltonian 
$H_{TLL}$~\cite{FurusakiZhang}, so long as $J_\perp\ll J$. The bosonization of the $S^z$ operator takes the same shape as \eqref{SzBoso}, however with different numerical values for the Luttinger parameter $K$ (and $u$). 


\textit{Conformal anomaly in thermal spin correlators}.---
As explained in the introduction, the anomaly resides with the energy-momentum tensor, which in the bosonic TLL language is given by 
\begin{equation}\label{Tboson}
 \frac{T(w)}{2\pi u\hbar}
 =-\lim_{w'\to w}\left[\p\varphi(w')\p\varphi(w)+\frac{1}{4\pi(w-w')^2}\right]
\end{equation}
where $\p=\p_w$. A similar expression holds for $\bT(\bw)$.
This makes clear, together with \eqref{TcylinderOnePt}, that the thermal correlator \footnote{Readers unfamiliar with low-dimensional QFT are referred to the supplemental material \cite{appendix}, where the tools to derive \eqref{corrExpansion} are presented.}
\begin{equation}
 \corr{\p\varphi(w)\p\varphi(w')}_{\beta_0}
 %
 \overset{w'\to w}{=}-\frac{1}{4\pi(w-w')^2}-\frac{\pi}{12(u\hbar\beta_0)^2}
 \label{corrExpansion}
\end{equation}
contains the conformal anomaly. Indeed, plugging \eqref{corrExpansion} 
into \eqref{Tboson}, the thermal expectation value \eqref{TcylinderOnePt} is readily recovered, along with the well known values $\ca=1=\bca$ for the TLL. 
Since the the $S^z$ operator \eqref{SzBoso} contains $\p\varphi$, the conformal anomaly must reside in the Green's function~\footnote{This correlator is evaluated explicitly in the supplemental material \cite{appendix}.}
$G_{\beta_0}(\st_1,\st_2)
=
 \corr{S^z(\st_1)S^z(\st_2)}_{\beta_0}-\corr{S^z(\st_1)}_{\beta_0}\corr{S^z(\st_2)}_{\beta_0}$,
where we introduce the spacetime label $\st=(\tau,x)=(w,\bw)$ for observables containing both $w$ and $\bw$ dependence. In order to extract the conformal anomaly similar to \eqref{corrExpansion}, the coincidence limit $\st_2\to \st_1$ needs to be taken in $G_{\beta_0}(\st_1,\st_2)$, which is however divergent and a regularization is required. We choose to subtract the zero temperature analog, i.e. at $\beta_0\to\infty$, taken as well in the coincidence limit, providing the normalized squared variance of local magnetization,
\begin{align}\label{OnSiteVarianceDefinition}
 \cD(\st,\beta_0)=\lim_{\st'\to \st}\left[G_{\beta_0}(\st',\st)-G_{\infty}(\st',\st)\right]/S_\text{max}^2
\end{align}
where $S_\text{max}=\frac{\hbar}{2}$ is the magnitude of the spin. Note that $\cD$ is dimensionless. Careful evaluation of the limit yields~\footnote{Readers might be worried about the appearance of the lattice constant $a$ in this result, as it is a scale, which is not taken into account in the CFT language. The lattice constant appears here due to the bosonization prescription \eqref{SzBoso}. Basically, \eqref{SzBoso} remembers the lattice origin of the Hamiltonian \eqref{SpinChain}, as it rightfully should, since this is the bridge between the lattice theory and the continuous low-energy effective QFT.}
\begin{align}\label{OnSiteVariance}
 \cD(\st,\beta_0)
 =
 \frac{2Ka^2}{\pi^2u\hbar}\left[\corr{T(w)}_{\beta_0}+\corr{\bT(\bw)}_{\beta_0}\right]
\end{align}
where \eqref{TcylinderOnePt} remains to be plugged in, along with $\ca=\bca=1$.
For DIMPY ((C$_7$H$_{10}$N)$_2$CuBr$_4$), one has $a=7.51$\AA{}, the TLL parameters $B\approx8.7$ Tesla to be $K\approx 1.2$, $u\hbar/a \approx2.34$meV \cite{Jeong}. This yields $\cD=1.1\times10^{-3}(\temp/\Kel)^2$, where $\Kel$ stands for Kelvin (and $\temp$ for temperature). 
At $B\approx20$ Tesla, one has $K\approx 1.2$, $u\hbar/a \approx1.62$meV, giving $\cD=2.3\times10^{-3}(\temp/\Kel)^2$. For copper pyrazine dinitrate (CuPzN) (Heisenberg spin chain) \eqref{OnSiteVariance} \cite{Hammar, breunig}, one has $a=6.7$\AA{}, $J=0.9$meV and TLL parameters $K=1/2$, $\hbar u=J a \pi/2$. This yields $\cD=1.2*10^{-3}(\temp/\Kel)^2$. The experimental detection of $\cD(\st,\beta_0)$ is discussed below; see Eqs. \eqref{integratedGamma}, \eqref{integratedSigma}.
\begin{figure}
%
%
\includegraphics[scale=.32]{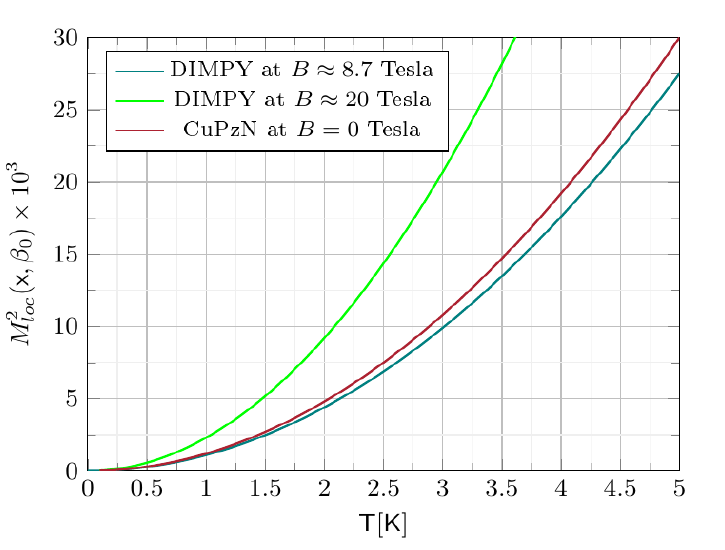}
\caption{Normalized squared variance of local magnetization $\cD(\st,\beta_0)$ for DIMPY and CuPzN.}
\label{figDplots}
\end{figure}

\textit{Temperature gradients and $\cD$
}.---
\begin{figure}
\includegraphics[scale=.32]{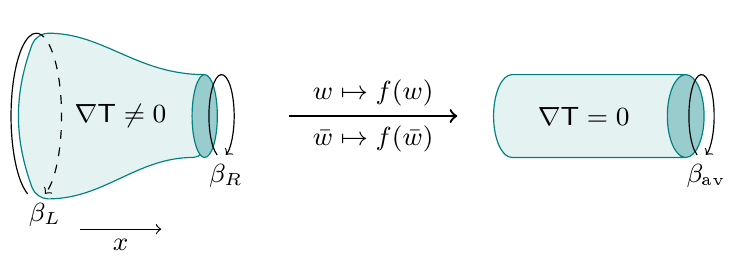}
\caption{In the Euclidean setting, a smooth, non-constant temperature profile may be seen as deformed cylinder with two distinct circumferences $\beta_{L,R}$ at its ends. This can be mapped into a system of constant average inverse temperature $\bbeta$ by the conformal transformation $y_\pm=f(w_\pm)=\int_0^{w_\pm}\bbeta/\beta(s)ds$. }
\label{figNonConstantTemp}
\end{figure}
From now on we revert to real time, $\tau=it$ so that $w=i(ut-x)$ and rotate this onto the real line, $w_-=iw$ and $w_+=-i\bw$, resulting in light-cone coordinates $w_\pm=x\pm ut$.

The signal \eqref{OnSiteVariance} can be amplified by preparing the system in a non-equilibrium initial state defined by a non-constant temperature profile $\beta(x)$ as discussed in~\cite{gawekdzki2018finite, langmann}, for instance
$\beta_{1}(x)
 =
 \frac{1}{2}\left(\beta_L + \beta_R - (\beta_L - \beta_R) \tanh\left[x/\delta\right]\right)$.
Any viable $\beta(x)$ interpolates smoothly between constant values $\beta_{L,R}=(k_B\temp_{L,R})^{-1}$ at the left and right extremes of the sample, with a kink at $x=0$, and $\delta$ indicating the rate of temperature change across the sample. This is depicted as deformed cylinder in \figref{figNonConstantTemp}. 
A conformal mapping $y_\pm=f(w_\pm)=\int_0^{w_\pm}\bbeta/\beta(s)ds$
reshapes this cylinder into one with uniform radius $\bbeta$, corresponding to a system of constant average temperature $(k_B\bbeta)^{-1}=(\temp_L+\temp_R)/2$. The new coordinates $y_\pm$ are naturally adapted to the symmetries of the system, similar to how spherical problems are best handled in polar coordinates. Indeed, the transformation \eqref{uniformizingTemp} shows that the deformed cylinder has the same amount of symmetries as the uniform temperature case, described in the previous section. 
As derived in \cite{gawekdzki2018finite}, the expectation value of the energy-momentum tensor becomes,
\begin{align}\label{TdeformedCylinderOnePt}
 \corr{T_\pm(w_\pm)}_{\beta(\st)}&=\frac{\pi^2\,\ca}{6u\hbar\beta(w_\pm)^2}-\frac{u\hbar\,\ca}{12}Sf(w_\pm),
%
\end{align}
with Schwarzian derivative $Sf(s)=\frac{1}{2}\left(\beta'(s)/\beta(s)\right)^2-\beta''(s)/\beta(s)$.

\begin{figure}
 \includegraphics[scale=.42]{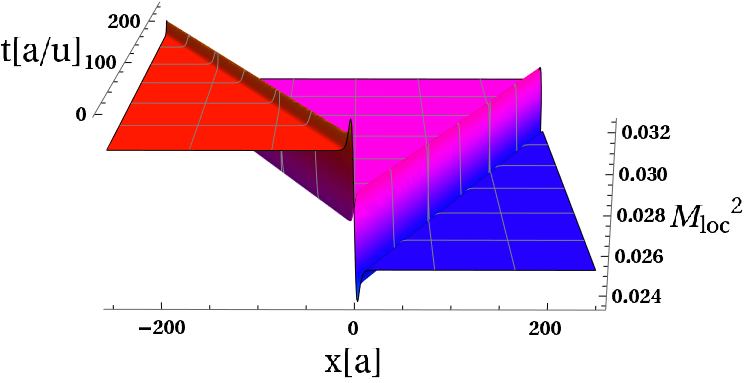}
 \caption{$\cD(\st,\beta(\st))$ from \eqref{OnSite_TempProfile} is depicted for temperature profile $\beta_1(x)$ in CuPzN over a region $x\in[-\ell,\ell]$ with $\ell=250a$ with $a=6.7$\AA{}. The parameters are $\temp_L=5\Kel$ (hot), $\temp_R=4.5\Kel$ (cold), $\delta=4a$. Time is measured in units $a/u\approx 0.5\,ps$.}
 \label{figDgradient}
\end{figure}

As explained in the supplemental material~\cite{appendix}, the derivation of $\cD$ is entirely analogous to the case of constant temperature and thus $\cD(\st,\beta(\st))$ is obtained by simply plugging \eqref{TdeformedCylinderOnePt} instead of \eqref{TcylinderOnePt} into \eqref{OnSiteVariance},
\begin{align}\label{OnSite_TempProfile}
 \cD(\st,\beta(\st))
 =
 \frac{2Ka^2}{\pi^2\hbar u}\sum_{\sigma=\pm}
 \biggl[\frac{\pi^2\,\ca}{6u\hbar\beta(w_\sigma)^2}-\frac{u\hbar\,\ca}{12}Sf(w_\sigma)\biggr]
 \,.
\end{align}
This shows that the lack of translation invariance in the spatial direction due to $\beta(x)$, also induces time dependence in $\cD$, as visualized in \figref{figDgradient} for CuPzN.
At the hot (cold) extreme of the sample, depicted in red (blue), $\cD(\st,\beta(x))=\cD(\st,\beta_{L\,(R)})$. At $t>0$ the peak separates in two waves which leave behind an equilibrated region (pink) for which $\cD(\st,\beta(\st))=\cD(\st,\bbeta)$.
$\cD$ is strongest in the vicinity of $x=0$ when $t=0$, where the magnitude of $\cD(\st,\beta(\st))>\cD(\st,\beta_L)$. 
We predict non-equilibium dynamics on the scale of nanoseconds (see \figref{figDgradient}), which should be measurable by ultrafast spectroscopy techniques \cite{lee2006ultrafast}. 
Experimental setups are discussed in the next section.

\textit{Experimental Consequences}.---
Focussing on constant temperature for now, we denote the space and time Fourier transformation of $G_{\beta_0}(x,t)$, i.e. the spin susceptibility, by $Q(k,\omega)$.
According to eq. \eqref{OnSiteVarianceDefinition},
%
%
$G_{\beta_0}(x,t)=\int d\omega \int dk Q(k,\omega) = g_0 + \cD(\beta_0;x,t)$, with $g_0$ a constant independent of temperature.
Therefore, a natural way to obtain $\ca$ from a spin system is to measure $Q(k,\omega)$ in neutron scattering experiments and integrate it in both momentum $k$ and energy $\hbar\omega$.
However, such a procedure is prone to difficulties in disentangling contributions to total structure factor coming from different scattering processes (both coherent, and incoherent) manifesting themselves in different regions of energy-momentum phase space. 
In addition, the precision of evaluation of $\ca$, or $\bca$, via $\int d\omega \int dk Q(k,\omega)$, in neutron scattering experiment might be scarcely limited by the coverage of $k$ and $\hbar\omega$ (\textit{i.e.} dynamic range) offered by the existing instruments.

Therefore, we suggest here two alternative ways of accessing the conformal anomaly in spin systems. For example, one can measure the local fluctuations $G(0,t)$ by NMR with a local probe \cite{herold2019dynamic, capponi2019nmr}, 
AFM \cite{schwarz2008magnetic} or $\mu$SR \cite{nuccio2014muon}
to obtain the \textit{local dynamical structure factor} $\Gamma(\omega)=\int G_{\beta_0}(x,t;x,0)e^{i\omega t}dt$.
%
%
It is related to $Q(k,\omega)$ by a momentum integration $\Gamma(\omega)=\int dk Q(k,\omega)$.
%
%
Instead of taking the detailed information of $Q(k,\omega)$,
the momentum-integrated quantity $\Gamma(\omega)$ is extracted
from experiment. Integrating it by $\omega$, one obtains the
conformal anomaly:
\begin{eqnarray}\label{integratedGamma}
\int_0^\infty d\omega \Gamma(\omega) \sim \ca \temp^2
\end{eqnarray}

This approach works well, as long as the measurement can access the vicinity of $\omega=0$. 
Otherwise, one needs another method to obtain $\Gamma(0)$, for instance, elastic neutron scattering, because it probes the information of $Q(k,0)$.

Another way to detect the conformal anomaly is to measure
the equal-time correlation by neutron diffraction, where one can obtain the energy-integrated momentum dependent structure factor $S(k)$. It produces the quantity $\Sigma(k)=\int Q(k,\omega)d\omega$.
%
%
By integrating over $k$,
\begin{eqnarray}\label{integratedSigma}
\int_{-\infty}^{+\infty} dk \Sigma(k) \sim \ca \temp^2
\end{eqnarray}
produces the conformal anomaly.
Furthermore, the information of $\Sigma(k)$ around $k=0$
can be reached by small-angle neutron scattering or by
noise measurement. How the latter relates to $Q(0,\omega)$
is explaiend as follows.

The average spin along the $z$-axis is denoted by
$\mathcal{S}=(\sum_i S^z_i /N) $ with $i$
the site index of the spins and $N$ the
total number of the spins. The noise measures the correlation function $\corr{\mathcal{S}(t)\mathcal{S}(0)}$ between time $t$ and 0 \cite{maeda2006noise, roy2015}. Noticing that
\begin{eqnarray}
\frac{1}{N^2} \sum_{i,j} \corr{S^z_i(t)S^z_j(0)}= \frac{1}{L} \int dx G(x,t),
\end{eqnarray}
one observe that the correlation function
$\corr{\mathcal{S}(t)\mathcal{S}(0)}$ measures
$Q(0,\omega)$ \cite{appendix}.

Now we turn to non-uniform (but static) temperature profiles.
Both ends of the spin chain are in contact with a heat reservoir of different temperature, $\temp_L$ and $\temp_R$. After a while, the system reaches a steady state,
such that the temperature profile does not change significantly with time.
At this time, we call it $t=0$, a non-trivial temperature profile $\temp(x)=(k_B\beta(x))^{-1}$ required for the discussion of the previous section arises.
The variance of local magnetization becomes \eqref{OnSite_TempProfile}, however, with $t=0$ fixed, and remains fixed for as long as the profile is stable. 

In order to detect the space dependent effects, one can perform the local measurement described 
above to obtain the local dynamical structure factor, which leads to Eq.\ref{integratedGamma}
but the temperature $\temp$ in the RHS will be replaced by 
the temperature profile in the final (steady) state $\temp(x)$. 
One can also measure the equal-time structure factor along the whole sample, 
e.g. by neutron diffraction to obtain some similar result as in Eq.\eqref{integratedSigma}.

Next, we turn to the case $t>0$. This corresonds to a situation where an initial state with profile $\beta(x)$ is reached and subsequently the heaters are turned off, so that the local magnetization equilibrates. 
In \figref{figDgradient}, this occurs within $\ell/u\sim$~ns where $\ell$ is the range of observation satisfying $\ell\gg\delta$.  Experimentally this can be investigated via ultrafast laser spectroscopy \cite{lee2006ultrafast} in the picosecond spectrum range (see our predictions in \figref{figDgradient}) via a pump-and-probe approach. 
This is reflected in optical properties such as reflectivity, absorption or Raman scattering. 


\textit{Conclusions}.---
In this paper, we have shown the presence of the conformal anomaly in thermal spin-spin correlators $G_{\beta}(\st_1,\st_2)$, concretely the normalized squared variance of local magnetization $\cD$. To our knowledge, this is the first instance of accessing the anomaly in a pure spin observable. We stress that our results are analytical even for strongly interacting $1d$ spin systems. Finally, we discussed experimental setups to probe the conformal anomaly. For a constant temperature profile, $\cD\propto \ca\temp^2$. The local magnetization can be measured in NMR, $\mu$SR and neutron diffraction as long as the massless Luttinger liquid is a good description of the spin system. 
Moreover, we showed how the application of temperature gradients across the sample introduces space and time dependence in the local magnetization $\cD$, see \eqref{OnSite_TempProfile}. In particular, the signal is amplified in a small neighborhood of the temperature kink at $x=0$. 
In spin chains with a steady but non-uniform temperature profile, one could measure a space-dependent local dynamical structure factor or could detect the magnetization dynamics in the ultrafast laser spectroscopy.

While we focussed on spin-$1/2$ chains and ladders, it would be interesting to see if our results can be extended to spin systems of spin $s>1/2$, which are associated with Wess-Zumino-Witten models \cite{affleck1989houches}. Another interesting possibilty is to extend our findings to higher order moments $\corr{S^z(0)\p_tS^z(\st)}$, which may be more accessible experimentally. Another direction is to consider space-dependent magnetic fields using methods from \cite{gawekdzki2018finite}.

\textit{Acknowledgements}.--- We acknowledge funding by the Deutsche Forschungsgemeinschaft (DFG, German Research Foundation) through SFB 1170, Project-ID 258499086, through the W\"urzburg-Dresden Cluster of Excellence on Complexity and Topology in Quantum Matter – ct.qmat (EXC2147, Project-ID 390858490) as well as by the ENB Graduate School on Topological Insulators.  We thank David Carpentier, Artem Odobesko, Anja Wenger for helpful discussions.

\newpage
\newpage

\begin{widetext}
\appendix
\section{The conformal Anomaly and the energy-momentum tensor}\label{appConfTransformations}
In CFT, there is a special class of fields, called primary fields, labelled by left- and right-moving conformal weights $(h,\bh)$, and characterized by their transformation behavior under conformal transformations $z\to g(z)$,
\begin{equation}\label{PrimaryTransformation}
 \chi_{h,\bh}\bigl(g(z),\bg(\bz)\bigr)
 =
 \left(\frac{\p g}{\p z}\right)^{-h}
 \left(\frac{\p \bg}{\p \bz}\right)^{-\bh}
 \,
 \chi_{h,\bh}(z,\bz)
\end{equation}
where $g(z)$ is a holomorphic function of $z$. In a free boson theory, such as the Luttinger liquid, one can easily write down two such fields. Starting from $\varphi(z,\bz)=\phi(z)+\bphi(\bz)$, the derivatives $\p\phi$ and $\bp\bphi$ are primary with conformal weights $(h,\bh)=(1,0)$ and $(h,\bh)=(0,1)$, respectively. 

The left-moving energy-momentum tensor $T(z)$ has dimensions $(h,\bh)=(2,0)$ while the right-moving component $\bT(\bz)$ has dimensions $(h,\bh)=(0,2)$. Due to the conformal anomaly, both fail to transform according to \eqref{PrimaryTransformation},
\begin{align}
 T(g(z))=\left(\frac{\p g}{\p z}\right)^{-2}\left[T(z)-\frac{\ca}{12}(Sg)(z)\right]\label{Ttransformation},\qquad
 \bT(\bg(\bz))=\left(\frac{\p \bg}{\p \bz}\right)^{-2}\left[\bT(\bz)-\frac{\bca}{12}(S\bg)(\bz)\right]
\end{align}
with Schwarzian derivative
\begin{equation}\label{Schwarzian}
 (Sg)(z)=\frac{g'''(z)}{g'(z)}-\frac{3}{2}\left(\frac{g''(z)}{g'(z)}\right)^2\,.
\end{equation}
A transformation from a zero-temperature system, coordinatized by $z$, to a thermal ensemble with temperature $1/\beta_0$, coordinatized by $w$, is
\begin{equation}\label{transformationCylinderApp}
 z=\exp\left(\frac{2\pi i w}{u\hbar\beta_0}\right),
 \qquad
 w=\frac{u\hbar\beta_0}{2\pi i}\log z\,,
 \qquad
 \bz=\exp\left(-\frac{2\pi i \bw}{u\hbar\beta_0}\right),
 \qquad
 \bw=-\frac{u\hbar\beta_0}{2\pi i}\log \bz\,.
\end{equation}
It has Schwarzian derivative $(Sw)(z)=\frac{1}{2z^2}$, so that 
\begin{equation}\label{Tcylinder}
 T_{\beta_0}(w)=-\left(\frac{2\pi}{u\hbar\beta_0}\right)^2\left[T_{plane}(z)z^2-\frac{u \hbar\, \ca}{24}\right],
 \qquad
 \bT_{\beta_0}(\bw)=-\left(\frac{2\pi}{u\hbar\beta_0}\right)^2\left[\bT_{plane}(\bz)\bz^2-\frac{u\hbar\,\bca}{24}\right]
\end{equation}
where we indicate the geometry on which the $T$ are defined by subscripts to be unambiguous. Note that we reinstated $u\hbar$ as prefactors of $\ca$ and $\bca$ for dimensional reasons. These expressions lead to a very simple evaluation of the expectation value of the energy-momentum tensor in a thermal ensemble. Indeed, fixing our reference energy such that $\corr{T_{plane}(z)}=0$, one easily finds
\begin{equation}
 \corr{T(w)}_{\beta_0}=\left(\frac{\pi}{\beta_0}\right)^2\frac{\ca}{6u\hbar}, 
 \qquad
 \corr{\bT(\bw)}_{\beta_0}=\left(\frac{\pi}{\beta_0}\right)^2\frac{\bca}{6u\hbar}
\end{equation}
where we have moved the subscript on $T_{\beta_0}$ to the expectation value in order to conform with the notation in the main text. Hence, the expectation value of the energy-momentum tensor is given entirely by the Schwarzian derivative term.

Since conformal transformations form a group the Schwarzian derivative does as well, meaning that concatenations of Schwarzian derivatives are again described by a Schwarzian derivative. This is relevant in turning to the non-constant temperature case described in the main text. There, the geometry of non-constant temperature $1/\beta(x)$ is first mapped into one with constant temperature and thereafter, similar to just now, into a zero-temperature system. Evaluating the Schwarzian derivative contribution in this scenario provides readily eqs. \eqref{TdeformedCylinderOnePt} where the Schwarzian derivative \eqref{Schwarzian} evaluated on
\begin{equation}\label{uniformizingTemp}
 y=f(w)=\int_0^{w}\frac{\beta_0}{\beta(\bs)}ds
\end{equation}
yields 
\begin{equation}
 Sf(s)=\frac{1}{2}\left(\frac{\beta'(s)}{\beta(s)}\right)^2-\frac{\beta''(s)}{\beta(s)}\,.\label{SchwarzianUniformization}
\end{equation}
as written in the main text. Note however that there we used Minkowskian coordinates and here, in the supplemental material, we stick with Euclidean coordinates. The energy-momentum tensor is then
\begin{align}
 T_{\beta(w)}(w)&=\left(\frac{\p f(w)}{\p w}\right)^2 T_{\beta_0}(f(w))+\frac{\ca}{12}Sf(w)\notag\\
 &=-\left(\frac{2\pi}{\beta(w)u\hbar}\right)^2 T_{plane}(z)\,z^2
 +
 \left(\frac{\pi}{\beta(w)}\right)^2\frac{\ca}{6u\hbar}
 +
 \frac{u\hbar\,\ca}{12}Sf(w)
\end{align}
and similarly for $\bT$. In going to the second line eq. $\eqref{Tcylinder}$ and $\frac{\p f}{\p w}=\frac{\beta_0}{\beta(w)}$ were employed. When taking the expectation value, $\corr{T_{plane}(z)}=0$ almost results in \eqref{TdeformedCylinderOnePt} after $w\to w_-$ and $T\to T_-$. We note here without derivation that the Schwarzian $Sf(w)$ picks up and extra sign when Wick rotating, $Sf(w)\to-Sf(w_-)$  This game is repeated for $\bT\to T_+$ with $\bw\to w_+$. This leads to \eqref{TdeformedCylinderOnePt}, which is derived entirely in real time in \cite{gawekdzki2018finite}. Our derivation here is the Euclidean time analog.

There is an important caveat when translating our Euclidean derivation to real time stemming from a subtlelty in Wick rotating, which is discussed at length in section 5 of \cite{gawekdzki2018finite}: The expectation value of the energy-momentum tensor picks up an additional Schwarzian-like contribution stemming from the curvature of the deformed cylinder in \figref{figNonConstantTemp} when moving between real and imaginary time. Working striclty in either real or imaginary time from the start, these contributions do not appear, as the derivations here and in \cite{gawekdzki2018finite} show. Since we are interested in settings in Minkowski spacetime, we worked with real time in the main text from the start. For the purposes of this appendix we stick with Euclidean time, to show that everything may be computed directly either in real or imaginary time.

\section{Correlators}\label{appCorrelators}
Two-point functions at zero temperature are given by
\begin{equation}\label{CFTzeroT2pt}
 \corr{\chi_{h_1,\bh_1}(z_1,\bz_1)\,\chi_{h_2,\bh_2}(z_2,\bz_2)}=\cN\frac{\delta_{h_1,h_2}}{(z_1-z_2)^{2h_1}}\frac{\delta_{\bh_1,\bh_2}}{(\bz_1-\bz_2)^{2\bh_1}}
\end{equation}
The constant $\cN$ is a normalization for each operator and usually it is chosen to be one. Here, it is left undetermined, so that the reader may adapt this to his favorite conventions.

For primaries, correlators at finite temperature can easily be obtained from \eqref{CFTzeroT2pt} by using the transformation \eqref{PrimaryTransformation} together with the thermalizing conformal transformation \eqref{transformationCylinderApp}. This yields
\begin{equation}\label{2pt}
 \corr{\chi_1(w_1,\bw_1)\chi_2(w_2,\bw_2)}_{\beta_0}
 =
 \cN\left(\frac{\pi}{u\beta_0\hbar}\right)^{2(h_1+\bh_1)}\sin^{-2h_1}\left(\frac{\pi(w_1-w_2)}{u\beta_0\hbar}\right)\sin^{-2\bh_1}\left(\frac{\pi(\bw_1-\bw_2)}{u\beta_0\hbar}\right)\delta_{h_1,h_2}\delta_{\bh_1,\bh_2}
\end{equation}

This opens the door to the computation of $G_{\beta_0}(\st_1,\st_2)
=
 \corr{S^z(\st_1)S^z(\st_2)}_{\beta_0}-\corr{S^z(\st_1)}_{\beta_0}\corr{S^z(\st_2)}_{\beta_0}$ employed in the main text,
\begin{align}
 G_{\beta_0}(\st_1,\st_2)&=\corr{S^z(\st_1)S^z(\st_2)}_{\beta_0}-\corr{S^z(\st_1)}_{\beta_0}\corr{S^z(\st_2)}_{\beta_0}\\
 &=
 a^2\rhbar^2\corr{\frac{\p_x\varphi(\st_1)}{2\pi R}\frac{\p_x\varphi(\st_2)}{2\pi R}}_{\beta_0}\notag\\
 &\quad+
 a^2\rhbar^2\corr{\frac{1}{\pi}:\cos\left(\frac{\varphi(\st_1)}{R}-2k_F(h)x_1\right):\frac{1}{\pi}:\cos\left(\frac{\varphi(\st_2)}{R}-2k_F(h)x_2\right):}_{\beta_0}\\
 &=K\left(\frac{a}{2\beta_0 u }\right)^2\left[\sin^{-2}\left(\frac{\pi(w_1-w_2)}{u\beta_0\hbar}\right)+\sin^{-2}\left(\frac{\pi(\bw_1-\bw_2)}{u\beta_0\hbar}\right)\right]\label{varianceResult}\\
 &\quad 
 +\rhbar^2\frac{\lambda \,a^2}{(2\pi)^2}\left(\frac{\pi}{\beta_0u\hbar}\right)^{2K}\sin^{-K}\left(\frac{\pi(w_1-w_2)}{u\beta_0\hbar}\right)\sin^{-K}\left(\frac{\pi(\bw_1-\bw_2)}{u\beta_0\hbar}\right)2\cos(k_F(h)(x_1-x_2))\notag
\end{align}
with a model dependent constant $\lambda$. In going to the second line, \eqref{SzBoso} and $\corr{S^z(w_2)}_{\beta_0}=m$ was employed, together with the fact that none of the fields in $S^z$ have the same conformal weights $(h,\bh)$. Therefore, there are no cross correlators between fields; for instance $\corr{\p_x\varphi\,:\cos\left(\frac{\varphi(w_1)}{R}-2k_F(h)x_1\right):}=0$. Using
\begin{equation}
 \frac{\p_x\varphi(w)}{2\pi R}=\sqrt{K}(J(w)+\bJ(\bw))
\end{equation}
and the fact that $J=\frac{i}{\sqrt{\pi}}\p\varphi$ has $(h,\bh)=(1,0)$ and $\bJ=-\frac{i}{\sqrt{\pi}}\bp\varphi$ has $(h,\bh)=(0,1)$ as well as $h=\bh=K/2$ for the cosine operator, the result \eqref{varianceResult} is straightforwardly derived through application of \eqref{2pt}. The correlator of the cosine operators is best handled by writing them in terms of exponentials and using the global $U(1)$ symmetry of the model which constrains correlators according to $\corr{e^{i\alpha \varphi(z_1)}e^{i\beta \varphi(z_2)}}\propto\delta_{\alpha+\beta,0}$.

Using expansions as in \eqref{corrExpansion} on the full correlator \eqref{varianceResult}, it is possible to take the limit for the on-site variance \eqref{OnSiteVarianceDefinition},
\begin{align}
 \cD(\st,\beta_0)&=\lim_{\st'\to \st}\frac{1}{S_\text{max}^2}\left[\corr{S^z(\st)S^z(\st')}_{\beta_0}-\corr{S^z(\st)}_{\beta_0}\corr{S^z(\st')}_{\beta_0}-\left(\corr{S^z(\st)S^z(\st')}_{\infty}-\corr{S^z(\st)}_{\infty}\corr{S^z(\st')}_{\infty}\right)\right]\notag\\
 &=\lim_{\st'\to \st}\frac{\rhbar^2a^2}{S_\text{max}^2}\biggl[\frac{K}{(2\pi)^2}\frac{1}{(w-w')^2}+\frac{K}{(2\pi)^2}\frac{1}{(\bw-\bw')^2}+\frac{1}{(2\pi)^2}\frac{\lambda}{|w-w'|^{2K}}+\frac{K\ca}{12(u\beta_0\hbar)^2}+\frac{K\bca}{12(u\beta_0\hbar)^2}+\cO(w-w')\notag\\
 &\qquad\qquad
 -\left(\frac{K}{(2\pi)^2}\frac{1}{(w-w')^2}+\frac{K}{(2\pi)^2}\frac{1}{(\bw-\bw')^2}+\frac{\lambda}{(2\pi)^2}\frac{1}{|w-w'|^{2K}}\right)\biggr]\notag\\
 &=\frac{2K a^2}{u\hbar \pi^2}\left[\corr{T(w)}_{\beta_{0}}+\corr{\bT(\bw)}_{\beta_{0}}\right]
\end{align}

This analysis can be repeated for the case of non-constant temperature. First, a two point function is now found by the additional transformation \eqref{uniformizingTemp} relating constant and non-constant temperature,
\begin{align}
 \corr{\chi_1(w_1,\bw_1)\chi_2(w_2,\bw_2)}_{\beta}
 =
 \left(\frac{df}{dw}\right)^{h_1}\biggr|_{w=w_1}\left(\frac{df}{dw}\right)^{h_2}\biggr|_{w=w_2}\left(\frac{df}{d\bw}\right)^{\bh_1}\biggr|_{\bw=\bw_1}\left(\frac{df}{d\bw}\right)^{\bh_2}\biggr|_{\bw=\bw_2}
 \corr{\chi_1(f(w_1),f(\bw_1))\chi_2(f(w_2),f(\bw_2))}_{\beta_0}\notag\\
 =
 \cN\left(\frac{\pi^2}{u^2\hbar^2\beta(w_1)\beta(w_2)}\right)^{h_1}
 \left(\frac{\pi^2}{u^2\hbar^2\beta(\bw_1)\beta(\bw_2)}\right)^{\bh_1}
 \sin^{-2h_1}\left(\int_{w_2}^{w_1}\frac{\pi}{u\hbar\beta(s)}ds\right)\sin^{-2\bh_1}\left(\int_{\bw_2}^{\bw_1}\frac{\pi}{u\hbar\beta(\bar{s})}d\bar{s}\right)\delta_{h_1,h_2}\delta_{\bh_1,\bh_2}
\end{align}
This allows to compute the analog of \eqref{corrExpansion}
\begin{subequations}\label{PhiCorrNeq}
\begin{align}
 &\corr{\p\varphi(w_1)\p\varphi(w_2)}_{\beta}
 =
 -\left(\frac{\sqrt{\pi}}{2u\hbar}\right)^2\frac{1}{\beta(w_1)\beta(w_2)}\sin^{-2}\left(\int_{w_2}^{w_1}\frac{\pi}{u\hbar\beta(s)}ds\right)\label{thermalPhiCorrNeq}\\
 &\overset{w_1\to w_2}{=}-\frac{1}{4\pi(w_1-w_2)^2}-\left[\frac{\pi}{12(u\hbar\beta(w_2))^2}+\frac{1}{24\pi}\left(\frac{1}{2}\left(\frac{\beta'(w_2)}{\beta(w_2)}\right)-\frac{\beta''(w_2)}{\beta(w_2)}\right)\right]+\cO((w_1-w_2))\label{corrExpansionNeq}\\
&=-\frac{1}{4\pi(w_1-w_2)^2}-\left[\frac{\pi}{12(u\hbar\beta(w_2))^2}+\frac{1}{24\pi}Sf(w_2)\right]+\cO((w_1-w_2))
\end{align}
\end{subequations}
Note again that, when Wick rotating, $Sf(w)\to -Sf(w_-)$.
In contrast to the case of constant temperature, the Schwarzian derivative \eqref{SchwarzianUniformization} appears. The other difference is of course the replacement $\beta_0\to\beta(w_2)$ in the first piece of the $\cO((w_1-w_2)^0)$ term. As before, by adjusting proportionality factors, the $\cO((w_1-w_2)^0)$ term is identified with $\corr{T}_\beta$, found in \eqref{TdeformedCylinderOnePt}. Then, by the same logic that led up to \eqref{OnSiteVariance} one can compute $\cD(\st,\beta(\st))$. The analysis is identical, so one is free to simply plug \eqref{TdeformedCylinderOnePt} into \eqref{OnSiteVariance}, as mentioned in the main text.

\section{About the upper bound of the energy/momentum measurement.}
In reality, measuring $Q(k,\omega)$ or $\Gamma(\omega)$ at
$\omega=\infty$ is impractical. One has to introduce a cut-off
energy $\omega_c$ as a truncation. How do we choose $\omega_c$
to avoid huge errors when we try to measure the conformal anomaly?
To answer this question, it is necessary to analyze
the two-point Green function
\begin{eqnarray}
G_{\beta}(x,t)=
\frac{{\rm Tr} \Big( e^{-\beta H}\,\p_x\phi(x,t)\,\p_x\phi(0,0) \Big)}
 {{\rm Tr} (e^{-\beta H})}.
\end{eqnarray}
With the help of mode expansion
\begin{eqnarray}
\phi(x,t)= \int \frac{dp}{2\pi \sqrt{2\epsilon_p}}
 \Big( a_p e^{-i\epsilon_p t +ipx} + a^{\dagger}_p e^{i\epsilon_p t -ipx} \Big),
\end{eqnarray}
we find that $G_{\beta}(x,t)=g(x,t)+\bar{G}_{\beta}(x,t)$, in which
\begin{eqnarray}
g(x,t)&=& \int \frac{dp}{4\pi \epsilon_p} p^2 e^{-i\epsilon_p t +ipx}, \\
\bar{G}_{\beta}(x,t)&=& \int \frac{dp}{4\pi \epsilon_p} p^2 n_B(\epsilon_p) cos(\epsilon_p t - px).
\end{eqnarray}
Only the latter function, $\bar{G}_{\beta}(x,t)$, carries the full temperature dependence. 
Comparing with \eqref{corrExpansion}, it must contain the information of the conformal anomaly. 
One recognizes the bosonic distribution function $n_B(\epsilon_p)=1/(e^{\beta\epsilon_p}-1)$, 
which decreases quickly as $\epsilon_p \gg k_B T$. 
For example, if the experiments are carried out below 10$\temp$, then several meV's will be
enough for the energy range of the measurement.
Such a truncation has the added benefit, that for the spin ladder, which contains high-lying energy modes, their energy range is not accessed and thus one needs not worry about their details.

\end{widetext}
\bibliographystyle{apsrev4-2}
\bibliography{refs.bib}

\end{document}